# Synchronization of optomechanical cavities by mechanical interaction

M. F. Colombano[1,2,†], G. Arregui[1,2,†], N. E. Capuj[3,4], A. Pitanti[5], J. Maire[1], A. Griol[6], B. Garrido[7], A. Martinez[6], C. M. Sotomayor-Torres[1,8], D. Navarro-Urrios[7]*

[1]Catalan Institute of Nanoscience and Nanotechnology (ICN2), CSIC and BIST, Campus UAB, Bellaterra, 08193 Barcelona, Spain.

[2]Depto. Física, Universidad Autónoma de Barcelona, Bellaterra, 08193 Barcelona, Spain.

[3]Depto. Física, Universidad de La Laguna, 38200 San Cristóbal de La Laguna, Spain.

[4]Instituto Universitario de Materiales y Nanotecnología, Universidad de La Laguna, 38071 Santa Cruz de Tenerife, Spain.

[5]NEST, CNR - Istituto Nanoscienze and Scuola Normale Superiore, Piazza San Silvestro 12, 56127 Pisa, Italy

[6]Nanophotonics Technology Center, Universitat Politècnica de València, 46022 València, Spain.

[7]MIND-IN2UB, Departament d'Enginyeria Electrònica i Biomèdica, Facultat de Física, Universitat de Barcelona, Martí i Franquès 1, 08028 Barcelona, Spain

[8]Catalan Institute for Research and Advances Studies ICREA, 08010 Barcelona, Spain.

† These authors contributed equally to this work.

* Corresponding author. Email: dnavarro@ub.edu

The synchronization of coupled oscillators is a phenomenon found throughout nature. Mechanical oscillators are paradigmatic among such systems, but realising them at the nanoscale is challenging. We report synchronization of the mechanical dynamics of a pair of optomechanical crystal cavities that are intercoupled with a mechanical link and support independent optical modes. In this regime they oscillate in anti-phase, which is in agreement with the predictions of our numerical model that considers reactive coupling. Finally, we show how to temporarily disable synchronization of the coupled system by actuating one of the cavities with a heating laser, so that both cavities oscillate independently.

Our results can be upscaled to more than two cavities and are thus the first step towards realizing integrated networks of synchronized optomechanical oscillators. Such networks promise unparalleled performances for time-keeping and sensing purposes and unveil a new route for neuromorphic computing applications.

# I. INTRODUCTION

Synchronization of autonomous oscillators, first observed back in the 17[th] century by Lord Huygens [1], manifests itself throughout nature extending from subatomic to cosmic scales, covering widely different research topics from biology to astrophysics [2]-[5]. This explains the existence of a vast literature devoted to synchronized oscillator networks, which have attracted much interest for decades [7]-[9], in part due to their applicability in neural networks [10]. With the advent of nanotechnologies, large efforts have been dedicated to synchronizing oscillating nanoelectromechanical systems (NEMS), which have wide practical applications because of their scalable architecture and accurate control of operating frequencies and quality factors by design [11]-[13]. Self-driven synchronized NEMS oscillator networks will find a variety of additional applications, such as on-chip robust time keeping [14] and mass [15], gas [16],[17] and force sensors [18] with extremely low phase noise [2]. The impressive progress in cavity optomechanics during the last decade [19]-[23] has evidenced that optomechanical (OM) oscillators are also ideal building blocks for the observation, control and exploitation of synchronization phenomena [24]. However, the field of OM oscillator networks is still in its infancy, being restricted to purely theoretical work and experiment proposals [24],[25]. Essential conditions for spontaneous synchronization between two dynamical systems are [2]: i) both of them are self-sustained oscillators, i.e., capable of generating their own rhythms; ii) the systems adjust their rhythm due to a weak interaction; and iii) the adjustment of rhythms occurs in a certain range of mismatch of the individual systems. To date, there has been only a handful of reports claiming synchronization in coupled OM cavities [26]-[28]. Most of these works are rather controversial [29], since the systems operate in a rather strong-optical-coupling regime that makes them indecomposable. Indeed, the coupled OM cavities share a common optical mode, which in addition prevents extracting the dynamics of each cavity in an independent way. On the other hand, a couple of reports have reported long-range synchronization between OM cavities placed in different chips [30],[31]. The coupling mechanism in these cases relies on modulating the optical excitation of one of the cavities with an electro-optical-modulator that reproduces the dynamics of the other cavity. Concerning NEMS with purely mechanical coupling, Shim et al. [11] reported a resonant excitation of coupled resonators using an external source, i.e., the oscillators were not self-sustained (aside from being strongly coupled).

In this Article, we unambiguously demonstrate spontaneous synchronization of the coherent mechanical oscillations of a pair of one-dimensional silicon OM photonic crystals (OMCs) integrated in the same chip that weakly interact by means of an engineered mechanical link, thus avoiding the need for external feedback loop schemes. The OMCs, which are optically isolated from each other, are independently driven to a state of high-amplitude, coherent and self-sustained mechanical motion (mechanical lasing from now on) using a self-pulsing (SP) dynamics [32],[33]. Indeed, the SP dynamics induces an anharmonic modulation of the optical force that can drive a mechanical mode displaying a significant single-particle OM coupling rate ($g_o$) into the mechanical lasing regime using either the fundamental or higher frequency harmonics of the force [32].

## II. SAMPLE DESCRIPTION AND EXPERIMENTAL SETUP

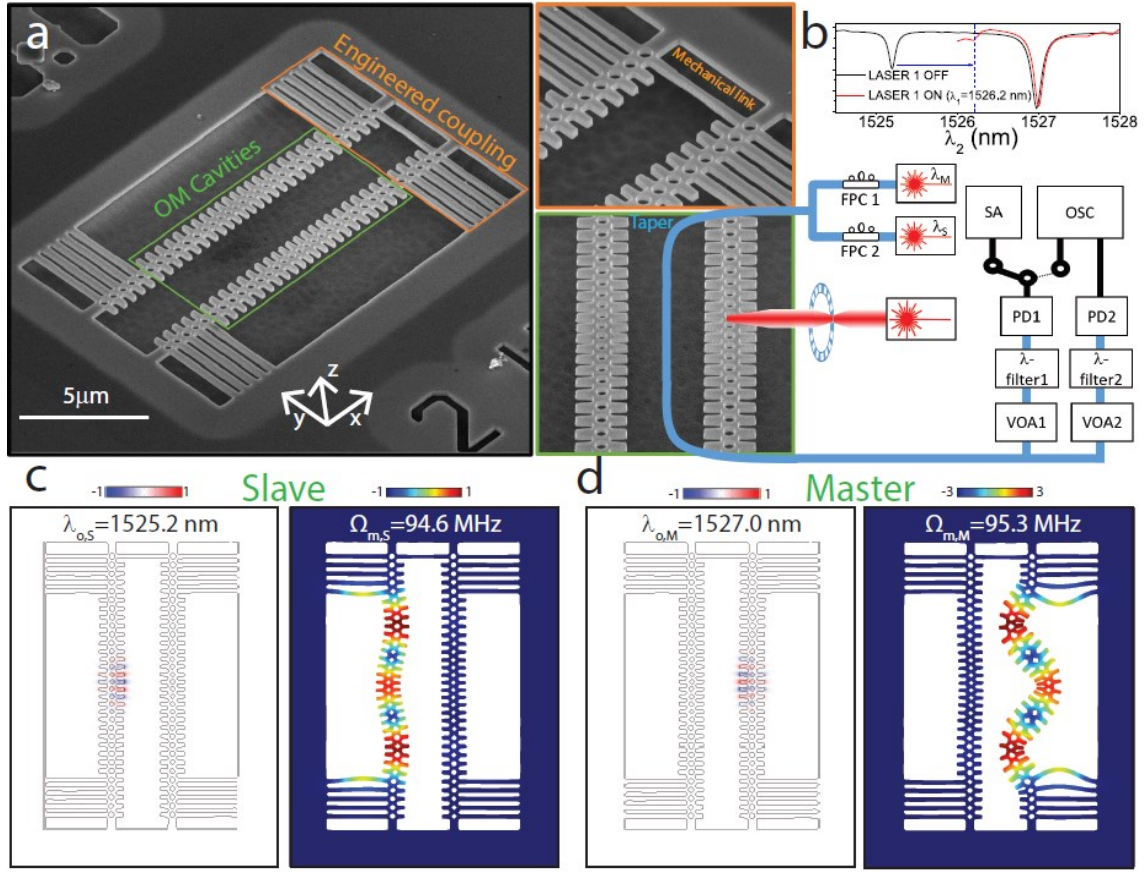

**FIG. 1. Mechanically-coupled Optomechanical Crystal Cavities. a,** SEM image of the pair of OMCs under study. **b,** Schematic of the experimental set-up. The optical transmission of a single cavity can be isolated from the other by means of λ-filters. VOA, variable optical attenuator; FPC, fibre polarizer controller; SA, spectrum analyser; OSC, oscilloscope; PD, photodetector. The top subpanel shows a transmission spectrum displaying resonances belonging to each cavity (black) and a spectrum of the longer-wavelength resonance while redshifting the first (red). **c** and **d,** FEM simulations of the normalized optical $E_y$ field and the mechanical displacement field $|Q|$ of the optical and mechanical modes under study using an imported geometry measured by SEM. The eigenmodes associated to the left (Slave) and right (Master) OMCs are depicted in panels c) and d), respectively. The displacement profiles have been altered to illustrate the larger oscillation amplitude of the Master.

The device investigated here is a pair of nominally-equivalent one-dimensional OMCs fabricated using standard Si nanofabrication processes (see Supplementary Materials) in a silicon-on-insulator wafer (Fig. 1a). The five outer cells at each side of the crystals are clamped to the partially underetched Si frame, so that the in-plane flexural modes are decoupled from the frame and confined to the central region of the OMCs, which are designed to support high-Q optical cavity modes at around 196 THz. The OMCs are mechanically interconnected on one side by a tether linking the inner stubs of the last cells. The physical separation between the crystals (≈2μm) is large enough to prevent optical crosstalk whilst allowing for their simultaneous optical excitation using a single tapered fibre that is placed in between.

In order to check whether the observed optical resonances belong to different OMCs we excite one of the resonances and redshift it using the thermo-optic (TO) effect while simultaneously monitoring the spectral position of the other resonance with the second laser (see the experimental setup on Fig. 1b). If only one resonance shifts it means that the two resonances belong to different OMCs, which is the case illustrated in the top panel of Fig. 1b.

We investigate the fundamental optical cavity mode (Fig. 1c) of each OMC. These modes display a slightly asymmetric field distribution with respect to the centre of the OMCs along the xz plane, giving rise to high values of $g_o$ for in-plane (xy plane) flexural modes. In particular, the ones having three antinodes along the x-direction (Fig. 1d) display a frequency of $\Omega_M$=95.3 MHz and $\Omega_S$=94.6 MHz in each OMC and a calculated value of $g_{o,M}/2\pi$=514 kHz and $g_{o,S}/2\pi$=330 kHz respectively (see Supplementary Materials). Hereinafter, we adopt the notation M and S to denote Master and Slave, where the OMC displaying higher $\Omega$ and longer optical resonance wavelengths is the Master and the other the Slave.

### III. SYNCHRONIZATION OF OPTOMECHANICAL OSCILLATORS IN THE FREQUENCY DOMAIN

The experiment that demonstrates synchronization of the dynamics of the two OMCs is illustrated in Fig. 2, where the left and right panels report the Radio-frequency (RF) spectra of the optical transmission associated to the laser exciting the Slave and Master, respectively. One laser is first switched on in between the two optical resonances and used to push the resonance of the Master to $\lambda_M$=1,531nm using the TO effect. The second laser is then used to excite the resonance of the Slave at $\lambda_S$=1,529nm, where it sets up a mechanical lasing regime at 94.6 MHz. A sharp and intense RF tone is observed when $\lambda$–filter1 is tuned at $\lambda_S$=1,529nm (Fig. 2a). It is worth noting that the configuration of the laser driving the Slave remains fixed for the rest of this particular experiment. The Master is mechanically excited by the coherent motion of the Slave as a consequence of the mechanical intercoupling of both OMCs. Indeed, when detecting at $\lambda_M$=1,531nm this excitation appears as a sharp RF tone at 94.6 MHz (Fig. 2b). A broad peak centred at 95.3 MHz is also present under this configuration, which is associated to the transduction of the thermally activated motion of the mechanical mode localized in the Master. By further red-shifting $\lambda_M$ up to $\lambda_M$=1,540.2 nm, a mechanical lasing regime is also established in the Master, where $\Omega_M$ decreases due to material heating. This latter effect appears to be also experienced by $\Omega_S$ to a weaker extent, i.e., part of the heat generated in the Master could end up increasing the effective temperature of the Slave. However, it is also plausible that the Master is frequency-pulling the Slave towards lower frequencies [34]. At this point, since both OMCs are in a mechanical lasing regime the spectral oscillation amplitude of the optical resonance is much larger than its linewidth and the transduction becomes extremely nonlinear [35]. Therefore, an additional RF peak appears at the beating frequency ($\Omega_M$-$\Omega_S$) together with symmetric sidebands at both sides of the lasing tones of the OMCs, which still lase at their own rhythm (Figs. 2c and 2d). The comparison of the RF spectra of both OMCs shows that the mechanical amplitude of the Master is much larger than that of the Slave as expected from the larger $g_{o,M}$ and $n_{o,M}$ values. The dynamical state of the two OMCs remain qualitatively the same until $\lambda_M$ reaches $\lambda_M$≈1,542.3nm, where the coupled system enters into a transition region. There, the Slave displays a complex RF spectrum of multiple peaks in which the main one is at $\Omega_M$ (see Figs. 2e and 2f, where $\lambda_M$≈1542.9nm). This state is not yet synchronization, even though the Slave is strongly affected by the dynamics of the Master. Synchronization is achieved above $\lambda_M \approx 1543$nm, where both OMCs coherently oscillate at $\Omega_{sync} \approx \Omega_M$ (see Figs. 2g and 2h, where $\lambda_M \approx 1543.6$nm). Both RF spectra look very similar with the remarkable difference of the presence of broad sidebands on the signal corresponding to the Slave, which are absent from the RF spectra of the Master. Those sidebands are clear signatures of Master-Slave synchronization and have been reported and analysed in previous works addressing synchronization of photonic cavities [28]. Their origin lies in the effect of the thermal force noise on the dynamics of the system. These forces push the Slave phase trajectory out from the limit cycle of the synchronized state, returning back in an oscillatory fashion at a frequency of ($\Omega_M$-$\Omega_S$). On the contrary, no sidebands appear on the Master spectrum, as it overdampedly returns to the limit cycle when driven away from it (see Supplementary Materials). The colour 2D plots of Figs. 2i and 2j reveal that the definition of the coupled system as a Master-Slave one is an oversimplification since $\Omega_M$ is also slightly pulled towards $\Omega_S$. This is evidenced in the abrupt frequency jumps of $\Omega_M$ both when entering the transition region above $\lambda_M \approx 1542.3$nm and the synchronized states above $\lambda_M \approx 1543$nm.

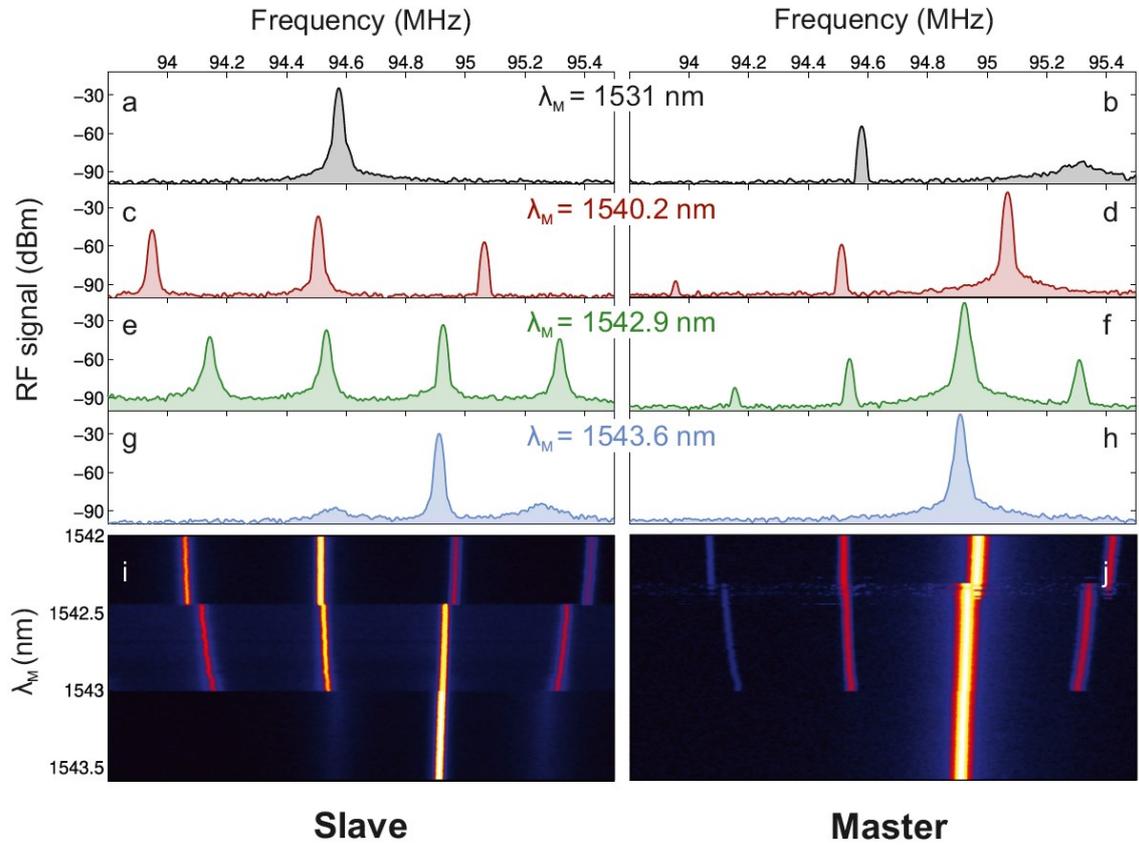

**FIG. 2. Synchronization of Optomechanical oscillators in the frequency domain. a-h,** Radio-frequency (RF spectra of the optical transmission associated to the Slave (left panels) and the Master (right panels) for different values of the wavelength of the laser driving the Master ($\lambda_M$). The wavelength of the laser driving the Slave is fixed at $\lambda_S$=1,529nm. **i-j,** Colour contour plots of the RF spectra as a function of $\lambda_M$ of the Slave and Master (i and j, respectively).

## IV. SYNCHRONIZATION OF OPTOMECHANICAL OSCILLATORS IN THE TEMPORAL DOMAIN

The temporal traces of the transmitted optical signal associated to each OMC are simultaneously recorded (Fig. 3) over a time span of 4 μs. On Fig. 3a we report the first

300 ns and a zoom of a single period appears in Fig. 3b. In order to analyse the quality of the synchronization signal, in Fig. 3c we show the Poincaré map associated to both temporal traces using a stroboscopic technique, i.e. we collect the pair of values {Transmission Master, Transmission Slave} at a specific sampling frequency. If the oscillators are synchronized and the temporal traces are sampled at $\Omega_{sync}$, a point on the Poincaré map is always found in the same position. In that case the phases of the two oscillators are the same for every sampled point of the traces, which is consistent with the experimental observations. Each of the coloured curves in Fig. 3c corresponds to a stroboscopic sampling at $\Omega_{sync}$ for a specific value of the initial delay ($\Delta t$, see Fig. 3b) and it is clear that the points remain in a confined volume of the phase space that is dominated by the experimental noise. By changing $\Delta t$ from 0 to $2\pi/\Omega_{sync}$ it is possible to rebuild the limit cycle of the synchronized state in the optical transmission phase space. An important point to consider here is that the maximum mechanical deformation of an OMC is achieved in-between the two minima of the optical transmission, as it is shown in Fig. 3b (see also Supplementary Materials). We can thus conclude that the temporal delay between the mechanical signals of each oscillator is about half of the period, i.e., there is a π phase shift between the mechanical oscillations.

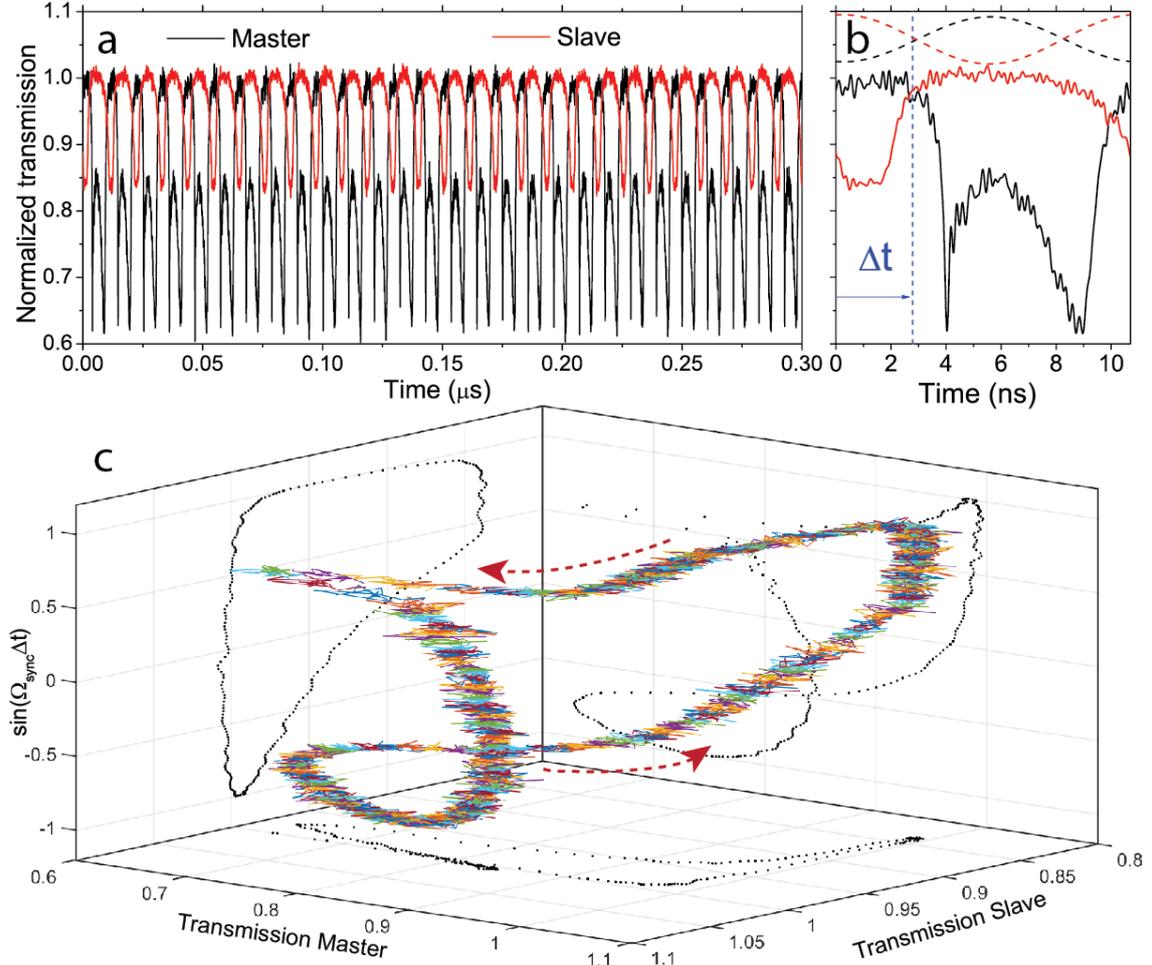

**FIG. 3. Temporal traces and Poincaré map of the synchronized state. a,** Temporal traces of the optical transmission of the Master and Slave cavities (black and red curves, respectively) as recorded simultaneously in the two channels of the oscilloscope. **b,** Zoom of a single period of the two signals. The simulated normalized generalized mechanical deformation corresponding to the experimental transmission traces are also reported (dashed lines). **c,** Poincaré map of the full temporal traces using the stroboscopic technique with a sampling frequency of $\Omega_{sync}$. Each coloured curve corresponds to a different value of the initial delay ($\Delta t$). The magnitude of the vertical axis has been chosen to be $sin(\Omega_{sync}\Delta t)$
to illustrate that the trajectory in the phase space is a closed cycle.

## V. NUMERICAL MODEL

To cast light on the physical mechanism behind the experimental results we implement a numerical model based on our previous studies on SP dynamics coupled to mechanical degrees of freedom through optical forces [32],[33]. It consists of a set of eight first order nonlinear equations, four for each OMC, describing the dynamics of the free carrier population ($N_i$), the average cavity temperature increase ($\Delta T_i$) and the generalized coordinates for the displacement of the mechanical modes and their derivatives ($u_i$ and $\dot{u}_i$, respectively), the subindex $i$ being $i$=M,S. For the sake of simplicity, the OMCs are considered equivalent in all their characteristic parameters but $\Omega_i$, which are taken from the experiment.
The specific characteristics of the coupling between the pair of OMCs are introduced in the harmonic oscillator equation of motion as a reactive (non-dissipative) contribution of the form $D(u_M-u_S)$, where $D$ is the coupling coefficient. This coupling term can be understood as an elastic restoring force caused by the deviation of the length of the linking tether from its rest value. In order to implement the Master-Slave configuration the coupling is considered to be unidirectional towards the Slave, i.e., the reactive term is only present in the Master harmonic oscillator equation. This situation is equivalent to considering bidirectional coupling with the Master oscillating with a much larger mechanical amplitude. The optical pumping parameters of the model are chosen such that a mechanical lasing regime is achieved in each OMC in the absence of coupling ($D$=0). More details on the numerical modelling can be found in the Supplementary Materials.
Figure 4 reports results of the stationary dynamics of the coupled system using a coupling coefficient that ensures achieving synchronization ($D$=1x10$^{15}$ s$^{-2}$) for decreasing values of $\Omega_M$, thus mimicking the heating effect when increasing $\lambda_M$ observed in the experiment. We define a relative change of $\Omega_M$ as $\Delta_M=(\Omega_M-\Omega_{M,o})/\Omega_{M,o}$, where $\Omega_M=\Omega_{M,o}$ at room temperature. The Fast Fourier Transform (FFT) of the simulated optical transmission associated to the Slave cavity as a function of $\Delta_M$ (Fig. 4a) shows several of the features appearing in the experimental results in Fig.2.
For $\Delta_M$=0 the main peak of the simulated RF spectrum of the Slave appears at a frequency $f\approx\Omega_S$ together with sidebands due to frequency beating. In this case, the phase diagram {$u_M$, $u_S$} almost fills the phase space in which the deformations are confined (Fig. 4b). By decreasing $\Delta_M$, the sidebands become closer to the main peak, which remains fixed. An abrupt transition occurs at $\Delta_M\approx$-2.1x10$^{-3}$, where the main Fourier peak switches to $f\approx\Omega_M$, entering a transition region that we associate to that observed in the experiment reported in Figs. 2e and 2f. Interestingly, by further decreasing $\Delta_M$ the relative dephasing between $u_M$ and $u_S$ becomes more and more confined (Fig. 4c). Mechanical synchronization occurs for $\Delta_M$<-4.5x10$^{-3}$, where a single peak appears at $f=\Omega_{sync}\approx\Omega_M$ with two symmetric sidebands that appear by introducing mechanical kicks in the $u_M$ and/or $u_S$ dynamical equations. Under those conditions, the FFT spectrum of the optical transmission associated to the Master cavity does not display sidebands (see Supplementary Materials). Moreover, as shown in Fig 4d, $u_M$ and $u_S$ appear to be $\pi$-shifted with respect to each other, which is consistent with the experimental observations in the synchronized state reported in Fig. 3.

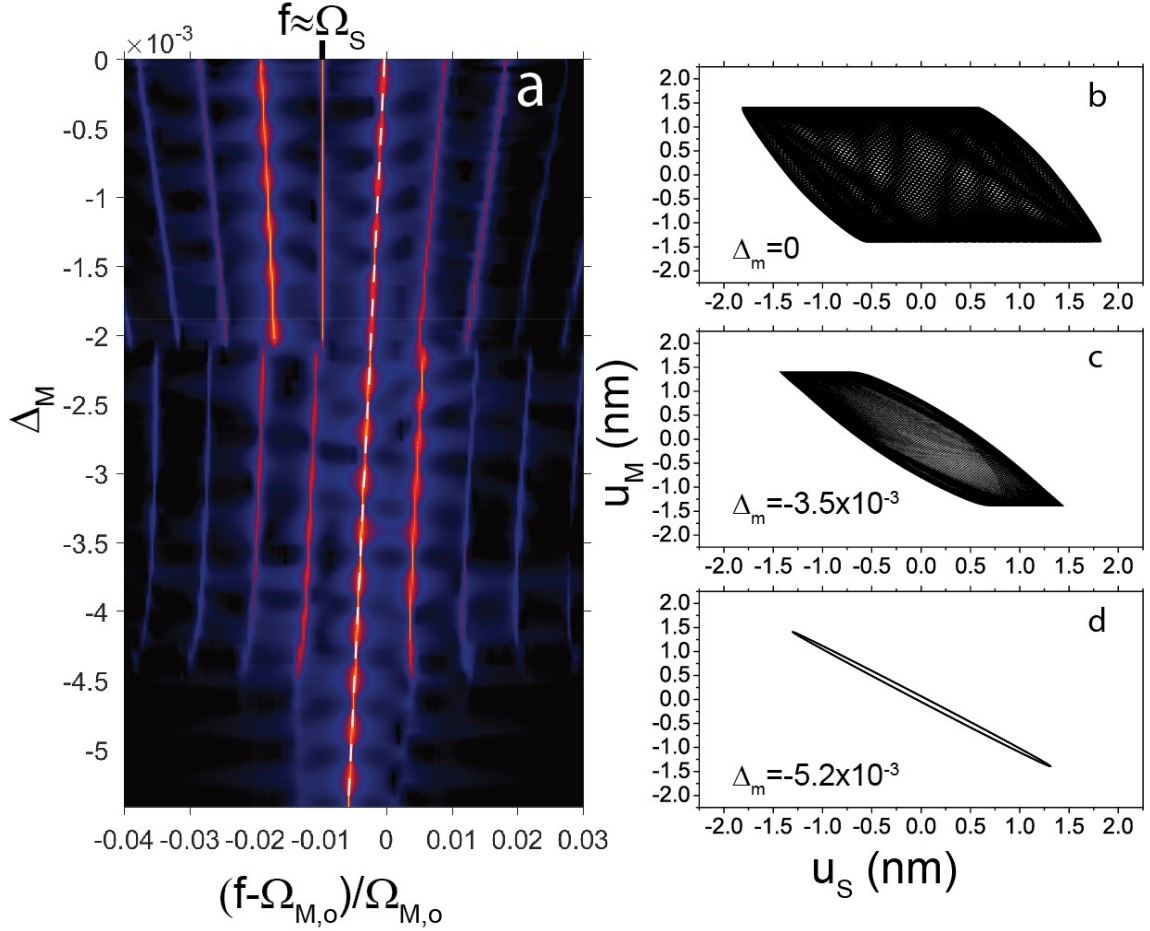

**FIG. 4. Numerical simulations of the coupled OMCs. a**, Colour contour plot of the simulated radio-frequency (RF) spectrum of the optical transmission of the Slave as a function of the normalized frequency shift of the Master $\Delta_M=(\Omega_M-\Omega_{M,o})/\Omega_{M,o}$. The frequency of the RF spectrum ($f$) is referred and normalized to the frequency of the Master at room temperature ($\Omega_{M,o}$). The dashed white line indicates the $f=\Omega_M$ curve **b-d**, Phase portraits of the deformation of the Master ($u_M$) and the Slave ($u_S$) for different values of $\Delta_M$.

## VI. ACTIVATION/DEACTIVATION OF THE SYNCHRONIZED STATE WITH AN EXTERNAL LASER SOURCE.

Finally, we explore the effect of illuminating the Master cavity region with a top heating laser. Our previous work on a single OMC showed that its dynamical state is modified when the laser is switched on because of photothermal effects [36]. Now, before switching on the top pumping, we set the parameters of the laser driving the Slave in a way that its dynamical state is a mechanical lasing regime at $\Omega_S$ activated by the 3rd harmonic of the optical force [32]. The Master is then driven to a standard mechanical lasing regime (using the 1st harmonic of the optical force), where both cavities synchronize their mechanical oscillations at $\Omega_{sync}$ (Figs. 5a and 5c). It is important to note that, under this configuration, the first harmonic of the optical transmission of the Slave is at $\Omega_{sync}/3$ and that there are no signs of the 1st and 2nd harmonic of that signal when measuring the Master. The latter observation is a conclusive evidence of a pure mechanical coupling between the OMCs, the leaked mechanical energy being enough to be transduced despite the rather low cross-coupling $g_o$.

When the top heating laser is switched on and the Master is illuminated the synchronized state is spoiled and a dynamical state similar to the one reported in Figs. 2c and 2d is achieved in the Slave (Fig. 5b). The Master still shows a coherent tone at $\Omega_M$ (Fig. 5d) but narrow symmetric sidebands appear associated to frequency beating with the coherent mechanical oscillation of the Slave. Although the Master is being heated when the top pump is on and hence its elastic constants are slightly relaxed, $\Omega_M$ shifts to a higher frequency value. This counterintuitive effect is a result of the attenuation of the frequency pulling effect induced by the coupling to the Slave mechanical dynamics, i.e., when the synchronized state is spoiled the coupling is reduced and $\Omega_M$ goes towards the eigenfrequency of the Master at that temperature. The same phenomenon occurs to $\Omega_S$ in the opposite direction and in a much larger scale (red arrow of Fig. 5b). By activating the modulation of the top pumping laser following the temporal profile of Fig. 5e it is possible to dynamically switch between the two states described above. The temporal traces when collecting the Slave and Master optical signals are reported in Figs. 5f and 5g, respectively. When the pump is switched off the OMCs take several microseconds to adjust their oscillation rhythms and stabilize the mechanical synchronized state, which is a direct consequence of their weak interaction.

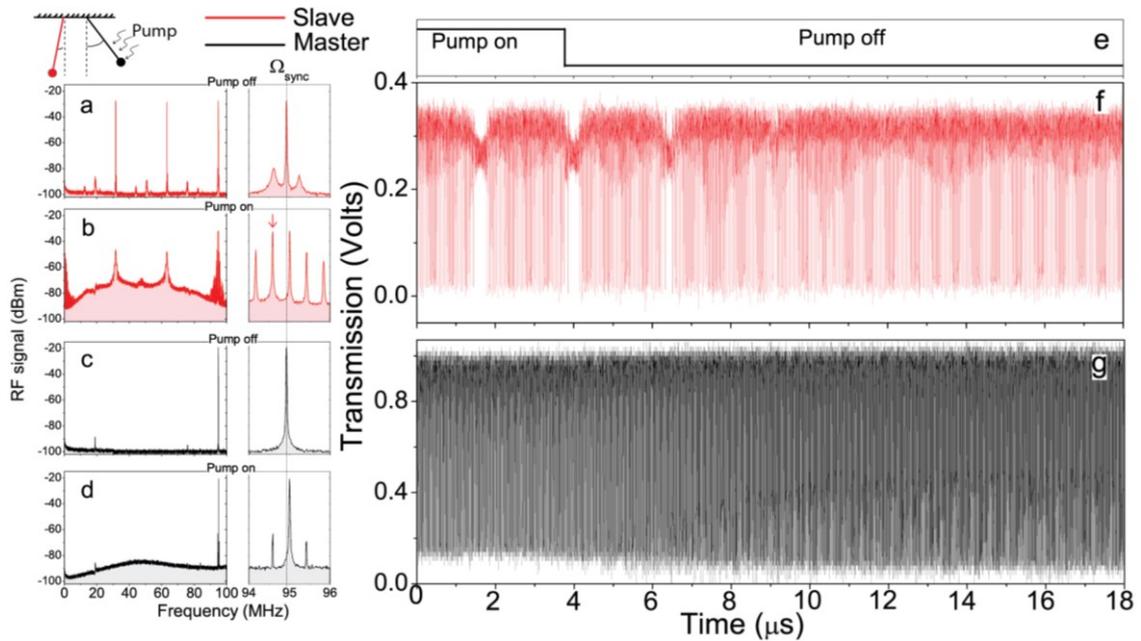

**FIG. 5. Switching on and off the synchronized state. a-d,** Radio-frequency spectra of the transmitted signal of the Slave (a and b) and Master (c and d) when the pump is off (a and c) and on (b and d). The right panels correspond to a zoom in the spectral region around the mechanical modes frequencies. The synchronized state frequency ($\Omega_{sync}$) is highlighted with a vertical line on the right panels. A vertical red arrow indicates $\Omega_M$ in the right panel b. **e,** Temporal trace of the top pumping laser when the modulation is active. **f-g,** Temporal traces of the optical transmission of the Master and Slave cavities (f and g respectively) when the Master is illuminated with a modulated top pumping laser.

## VII. CONCLUSIONS AND OUTLOOK

In conclusion, we have unequivocally demonstrated synchronization of the mechanical oscillations of a pair of optomechanical crystal cavities by introducing a weak mechanical coupling between them. Provided that our numerical model indicates that the observed features are compatible with a reactive type of coupling, further experimental studies will unveil whether the transition to synchronization is through phase locking or suppression of the natural dynamics, which are the two mechanisms of synchronization in reactively coupled oscillators [3]. Finally we have demonstrated that the synchronization dynamics can be switched back to an unsynchronized state by introducing an external heating source on the Master cavity as well as the possibility to dynamically modulate between those two states.

The system presented here, consisting in two parallel OMCs built on a silicon chip by standard nanofabrication processes, can be optically excited by a common laser source by exploiting thermo-optic effects while still maintaining the requirements for synchronization (see Supplementary Materials). Such a simple configuration could be easily upscaled to realize complex networks comprising many more nodes without substantially increasing the technological requirements. For instance, one could easily think of a silicon chip integrating an array of optomechanical cavities accessed optically via integrated waveguides [37],[38] and interconnected via mechanical links. Therefore, our results are the first step towards building networks of coupled optomechanical crystal cavities able to display collective dynamics prone to be modified by addressing single structures with external perturbations. These rather unique features are to be exploited in neuromorphic photonic computing applications [39], for instance for pattern recognition tasks or more complex cognitive processing. Eventually it will be possible to experimentally investigate the limits, in terms of network complexity, of coherent collective behaviours and the intriguing transition towards incoherence, where the so-called Chimera states [40],[41] of synchronous and incoherent behaviour are expected to emerge.

## ACKNOWLEDGMENTS

The authors thank Dr. G. Whitworth for his careful and critical reading of the manuscript. This work was supported by the European Commission project PHENOMEN (H2020-EU-713450), the Spanish Severo Ochoa Excellence program and the MINECO project PHENTOM (FIS2015-70862-P). DNU, GA and MFC gratefully acknowledge the support of a Ramón y Cajal postdoctoral fellowship (RYC-2014-15392), a BIST studentship and a Severo Ochoa studentship, respectively.

M.F.C., G.A. and D.N.-U. performed the experiments and analyzed the data. D.N.-U. and N.C. devised the experiment and developed the model. J.M. and A.P. participated in the experiments at different stages. A.G. and A.M. fabricated the samples. All authors contributed to the interpretation of the results and writing of the manuscript.

Authors declare no competing interests.

# Supplementary Materials for

## Synchronization of optomechanical cavities by mechanical interaction

M. F. Colombano, G. Arregui, N. E. Capuj, A. Pitanti, J. Maire, A. Griol, B. Garrido, A. Martinez, C. M. Sotomayor-Torres, D. Navarro-Urrios.

Correspondence to: dnavarro@ub.edu

**This PDF file includes:**

Materials and Methods
Figs. S1 to S7

## Materials and Methods

### Devices fabrication and design

The devices (Fig. 1a and Fig. S1) were fabricated in standard silicon-on-insulator (SOI) SOITEC wafers with silicon layer thickness of 220 nm (resistivity $\rho$ ~1-10 $\Omega$ cm$^{-1}$, p-doping of ~$10^{15}$ cm$^{-3}$) and a buried oxide layer thickness of 2 $\mu$m. The pattern was written by electron beam in a 100 nm thick poly-methyl-methacrylate (PMMA) resist film and transferred into silicon by Reactive Ion Etching (RIE). Application of BHF removed the buried oxide layer and released the beam structures.

The defect region of the OMCs consists of 12 central cells in which the pitch (a), the radius of the hole (r) and the stubs length (d) are decreased in a quadratic way towards the centre. The maximum reduction of the parameters is denoted by $\Gamma$. At both sides of the defect region a 5 period mirror followed by 5 cells clamped to the frame are included. The nominal geometrical values of the cells of the mirror are a=500nm, r=150nm, and d=250nm. The total number of cells is 32 and the whole device length is about 15$\mu$m. All the results presented in this work correspond to the structure with $\Gamma$=85%.

### Experimental setup

Light from two tuneable lasers is injected to a single tapered fibre to excite independently the fundamental optical resonance of each OMC, which appear at slightly different wavelengths due to the departure of the fabricated structures from their nominal parameters. The transmitted signals are then split into two optical paths and spectrally filtered so that each detector collects the signal from one of the lasers.

In the experiment, the optical resonance of the Master is characterized by an extrinsic and intrinsic radiative decay rates of $\kappa_{e,M}$=8x10$^{10}$ s$^{-1}$ and $\kappa_{i,M}$=2x10$^{10}$ s$^{-1}$, respectively and a number of intracavity photons in perfect resonance of $n_{o,M}$=4x10$^5$ ph. Concerning the Slave, the values are $\kappa_{e,S}$=0.6x10$^{10}$ s$^{-1}$, $\kappa_{i,S}$=5x10$^{10}$ s$^{-1}$ and $n_{o,S}$=1x10$^5$ ph.

The set-up also includes a top pumping scheme that allows to focus a modulated laser beam ($\lambda$=808nm) on top of one of the OMCs.

Temporal mismatches due to different optical and electrical delays between both paths of the experiment are taken into account by detecting simultaneously the master signal through both of the optical fiber channels of our setup.

All the measurements have been performed at atmospheric conditions of pressure and temperature.

### Single laser scheme

A synchronized state can also be achieved using a single laser scheme (not shown here). Indeed, since the TO wavelength shift can be much larger than the spectral difference between the resonances belonging to different OMCs, it is possible to simultaneously drive both OMCs to a mechanical lasing state and eventually to synchronization. However, the physical coupling mechanism in this configuration is probably more complex since, in addition to the mechanical interaction, a bidirectional optical coupling may be also playing a role, in a similar fashion to what was reported in Ref. *(41)* for cascaded OM disk resonators.

### FEM simulations and OM coupling calculations

To model accurately the fabricated pair of OMCs and account for the differences from the nominal design, the in-plane geometry was imported from the scanning electron microscopy (SEM) micrographs into the finite element method (FEM) solver (Fig. S1), where the thickness is that of the top Si layer of the SOI wafer. This procedure ensures a good agreement between the measured optical and mechanical modes and those extracted from simulations.

Among the many optical and mechanical modes supported by the OM crystal, we have considered those discussed along the main text, i.e., the fundamental optical modes and the in-plane flexural modes with three antinodes illustrated in Fig. 1. The calculated effective masses are $m_{eff}$=5 pg for both mechanical modes.

Single-particle OM coupling rates are estimated by taking into account both the photoelastic (PE) and the moving interfaces (MI) effects *(42-44)*. The PE effect is a result of the acoustic strain within bulk silicon while the MI mechanism comes from the dielectric permittivity variation at the boundaries associated with the deformation.

The calculation of the MI coupling coefficient $g_{MI}$ is performed using the integral given by Johnson et al. *(42)*;

$$g_{MI} = -\frac{\pi \lambda_r}{c} \frac{\oint (\mathbf{Q}\cdot\hat{\mathbf{n}})(\Delta\varepsilon |\mathbf{E}_\parallel|^2 - \Delta\varepsilon^{-1}|\mathbf{D}_\perp|^2)dS}{\int \mathbf{E}\cdot\mathbf{D}\,dV}\sqrt{\frac{\hbar}{2\Omega_m}} \tag{S1}$$

where **Q** is the normalized displacement (max{|**Q**|}=1), $\hat{\mathbf{n}}$ is the normal at the boundary (pointing outward), **E** is the electric field and **D** the electric displacement field. $\varepsilon$ is the dielectric permittivity, $\Delta\varepsilon = \varepsilon_{silicon} - \varepsilon_{air}$, $\Delta\varepsilon^{-1} = \varepsilon^{-1}_{silicon} - \varepsilon^{-1}_{air}$. $\lambda_r$ is the optical resonance wavelength, $c$ is the speed of light in vacuum, $\hbar$ is the reduced Planck constant and $\Omega_m$ is the mechanical mode eigenfrequency, so that $\sqrt{\frac{\hbar}{2\Omega_m}}$ is the zero-point motion of the resonator.

A similar result can be derived for the PE contribution *(43,44)*:

$$g_{PE} = -\frac{\pi \lambda_r}{c} \frac{\langle E|\delta\varepsilon|E\rangle}{\int \mathbf{E}\cdot\mathbf{D}\,dV}\sqrt{\frac{\hbar}{2\Omega_m}} \tag{S2}$$

where $\delta\varepsilon_{ij} = \varepsilon_{air} n^4 p_{ijkl} S_{kl}$, being $p_{ijkl}$ the PE tensor components, $n$ the refractive index of silicon, and $S_{kl}$ the strain tensor components.

The addition of both contributions results in the overall single-particle OM coupling rate:

$$g_o = g_{MI} + g_{PE} \tag{S3}$$

It is worth noting that the typical values of $g_{PE}$ for the string-like modes studied here are on the order of Hz, and therefore can be neglected in comparison to the MI counterpart.

In Fig. S2 we illustrate the MI surface density (the integrand of Eq. S1) associated to the Slave optical mode when evaluating the contribution of the Slave and the Master flexural modes (panels a and b respectively). Similar results are obtained when performing the same study for the Master optical mode. The slight asymmetry of the field distribution with respect to the centre of the OMCs along the xz plane unbalances the MI surface density towards the region of the OMC cavity that is closer to the centre of the frame, thus giving rise to overall calculated values of $g_{o,M}/2\pi = 514$ kHz and $g_{o,S}/2\pi = 330$ kHz for in-plane (xy plane) flexural modes belonging to the Master and Slave respectively.

Although the energy of the mechanical modes under consideration is mostly confined to the region between the clamping tethers of a single OMC, the presence of a mechanical link extends their spatial distribution to the central part of the other OMC. Cross contributions values of $g_{o,MS}/2\pi = 6$ kHz and $g_{o,SM}/2\pi = 8$ kHz were calculated as well, where the first subindex denotes the OMC associated to the optical mode. In this context, Fig. S2b shows, for the case of the flexural mode of the Master and the optical mode of the Slave, that the source of cross MI contribution is the small partial overlap of the mechanical deformation with the Slave Cavity region, while there is a negligible contribution due to the overlap of the Slave optical mode with the Master OMC. This confirms that the two OMCs can be considered to be optically isolated from each other by design.

## Numerical Model

In order to generate to generate a high amplitude, coherent and self-sustained mechanical motion (mechanical lasing from now on) in each of the OMCs, we exploit the so-called self-pulsing (SP), which is a limit-cycle solution that arises from the dynamical interplay between free-carrier-dispersion (FCD) and the thermo-optic (TO) effect. The SP mechanism has been reported before by our and other groups in Si based OMCs and a detailed description of the phenomenon can be found elsewhere *(31,45)*. As a result of the SP, the optical resonance oscillates periodically around the laser line, thus impinging an anharmonic modulation of the radiation pressure force within the cavity. If one of the frequency harmonics of the force is resonant with a mechanical mode displaying a significant $g_o$ value, that particular mode can be driven into the mechanical lasing regime *(31)*. The two OMCs are mechanically coupled by means of the linking tether.

The dynamics of the coupled pair of OMCs has been studied numerically by solving a model consisting of a set of eight first order nonlinear equations, four for each OMC, that describe the dynamics of: the free carrier population ($N_i$), the average cavity temperature increase ($\Delta T_i$) and the generalized coordinates for the displacement of the mechanical modes and their derivatives ($u_i$ and $\dot{u}_i$, respectively), the subindex i being i=M,S. For the sake of simplicity, we have considered that the OMCs are equivalent in all their characteristic parameters but the mechanical eigenfrequencies ($\Omega_i$), which have been taken from the experiment.

Thus, the full nonlinear system reads as follows:

$$\dot{N}_i = -\frac{1}{\tau_{FC}} N_i + \beta\left(\frac{hc^3}{n^2 \lambda_o V_o^2}\right) n_{o,i}^2 \qquad (S4a),$$

$$\dot{\Delta T_i} = -\frac{1}{\tau_T} \Delta T_i + \alpha_{FC} N n_{o,i} \qquad (S4b),$$

$$\ddot{u}_{m,i} + \cdots + \delta_{iS} D(u_M - u_S) = \frac{F_{o,i}}{m_{eff}} \qquad (S4c)$$

Equations S4a consider a Two-Photon Absorption (TPA) generation term, where $\beta$ is the tabulated TPA coefficient and a surface recombination term governed by a characteristic lifetime $\tau_{FC}$. $V_o$ is the optical mode volume and $\lambda_o$ is the cavity resonance wavelength at room temperature.

Equations S4b reflect the balance between the fraction of photons that are absorbed and transformed into heat due to free-carrier-absorption and the heat dissipated to the surroundings of the cavity volume, which is governed by a characteristic lifetime $\tau_T$. $\alpha_{FC}$ is defined as the rate of temperature increase per photon and unit free-carrier density.

Equations S4c are harmonic oscillator equations driven by optical forces ($F_{o,i}$). The specific characteristics of the mechanical coupling between the pair of OMCs have been introduced as a reactive (non-dissipative) contribution of the form $D(u_M - u_S)$, where D is the coupling coefficient. This coupling term can be understood as an elastic restoring force caused by the deviation of the length of the linking tether from its value at rest. In order to implement the Master-Slave configuration we have considered that the coupling is unidirectional towards the Slave, i.e., the reactive term is only present in the Slave harmonic oscillator equation. This situation is equivalent to considering bidirectional coupling with the Master oscillating with a much larger mechanical amplitude. The optical pumping parameters of the model are chosen so that a mechanical lasing regime is achieved in each OMCs in the absence of coupling (D=0).

A simple schematic of the coupled system is shown in Fig. S3. The two OMCs are coupled mechanically in a unidirectional way while, within the single OMC, the SP and the mechanics are coupled bidirectionally. The driving terms of Equations S4 depend on the number of intracavity photons ($n$), which can be written as $n = n_o \frac{\Delta\lambda_o^2}{4(\lambda_l - \lambda_r)^2 + \Delta\lambda_o^2}$, where $n = n_o = 2 P_l \kappa_e \lambda_o / \kappa^2 hc$ in perfect resonance. $P_l$ and $\lambda_l$ are the laser power and wavelength, respectively; $\kappa_e$ and $\kappa$ and are the extrinsic and overall optical damping rates, respectively, the latter determining the cavity resonant linewidth ($\Delta\lambda_o = \lambda_o^2 \kappa / 2\pi$). The position of the resonance incorporates the FCD and the TO contributions and the effect of the OM coupling with the mechanical mode localized in the same OMC and with that mostly localized in the other OMC. According to this, the optical resonance is written in the following terms:

$$\lambda_{r,i} \approx \lambda_{o,i} - \frac{\partial \lambda_{r,i}}{\partial N_i} N_i + \frac{\partial \lambda_{r,i}}{\partial T_i} \Delta T_i - \frac{\lambda_{o,i}^2 g_{o,i}}{2\pi c} u_i - \frac{\lambda_{o,i}^2 g_{o,ij}}{2\pi c} u_j$$

Importantly, the response of $n$ to the various contributions is adiabatic since $\kappa$ is much larger than $1/\tau_{FC}$, $1/\tau_T$ and $\Omega_{m,i}$.

The parameters governing equations S4a and S4b have been adjusted to reproduce the experimental SP dynamics when a mechanical mode is not being resonantly pumped. They are extracted from our previous works, $\tau_T$=0.5 [µs], $\tau_{FC}$=0.5 [ns] and $\alpha_{FC}$=4x10$^{-13}$ [K cm$^3$ s$^{-1}$], while the initial conditions verify that $\frac{\partial \lambda_{r,i}}{\partial \Delta T_i} \Delta T_i(t=0) = \lambda_{l,i} - \lambda_{o,i}$ and $N_i(t=0) = u_i(t=0) = \dot{u}_i(t=0) = 0$. TO and FCD coefficients were independently calculated by assuming that the observed wavelength shift is only associated to an average change in the Si refractive index within the region overlapping with the electromagnetic fields and using tabulated values for its dependence with T and N. This procedure leads to the following values: $\frac{\partial \lambda_r}{\partial N}$=7x10$^{-19}$ [nm cm$^3$] and $\frac{\partial \lambda_r}{\partial \Delta T}$=6x10$^{-2}$ [nm K$^{-1}$]. The simulations cover a temporal span of 1x10$^{-4}$ s discretized in time steps of 2x10$^{-11}$ s, which ensure both enough sampling rate to account for the fastest dynamics of the system and reaching a stationary dynamical regime.

Fig. S4 reports results of the stationary dynamics of the coupled system for different values of D. When analysing the Fast Fourier Transform (FFT) of the simulated optical transmission associated to the Slave cavity (Fig. S4a). In Fig. S4 we also analyse the relative dephasing of the mechanical deformations of each OMC by plotting $u_M$ with respect to $u_S$ (right panels of Fig. S4). As mentioned before, for D=0 both OMCs are in the phonon laser regime with no interaction, so the spectrum of Fig. S4a consists of a single peak at $\Omega_S$ and the phase diagram of Fig. S4b completely fills the phase space in which the deformations are confined. By increasing D, sidebands appear in the spectra of Fig. S4a, the relative dephasing becomes more and more confined (Supplementary Figs. 4d and 4e). Eventually the main Fourier peak switches to $\Omega_M$, which is the situation reported in Fig. S4e (D=8x10$^{14}$ s$^{-2}$). The dynamics of the two cavities become synchronized for D values greater than D~9x10$^{14}$ s$^{-2}$, where a single peak appears at $\Omega_{sync}$~$\Omega_M$ with two symmetric sidebands at a frequency of $\Omega_{sync} \pm (\Omega_M - \Omega_S)$.

In Fig. S5 we represent the stationary dynamics achieved in each OMC in a synchronized state obtained at D=10x10$^{14}$ s$^{-2}$ (these are the conditions used also for the results showed in Fig. 4) and the associated optical transmission temporal traces. It is worthwhile emphasizing that the deformation maximum is found in between the two minima of the transmission trace, i.e., where the limit cycle trajectory verifies $\lambda_l > \lambda_r$.

Thus, the direct OM coupling contribution not only locks the main frequency of the SP to a simple fraction of the mechanical eigenfrequency (in the case of Fig. S5 they are locked at the mechanical frequency of the Master) but also impacts the duty cycle by modifying the time distance between both transmission minima. The two trajectories belonging to Master and Slave appear to be $\pi$-shifted with respect to each other, which is consistent with the experimental observations in the synchronized state reported in Fig. 3. The simulated mechanical amplitudes are similar as we have assumed the two OMCs to be roughly equivalent. Thus, it is reasonable that both simulated temporal traces shapes look alike, in contrast to what observed experimentally. Indeed the experimental temporal trace of the optical transmission associated to the Slave (Figs 3a and 3b) shows a smaller duty cycle than the Master counterpart because the mechanical amplitude is sensibly smaller in the former case.

From the numerical simulations we also note that the effect of the cross-linked OM coupling is negligible in what concerns the overall dynamics of the system, obtaining equivalent results when $g_{o,MS}$=0. From this latter result it is possible to conclude that the meaningful coupling between the two OMCs is purely mechanical, i.e., the important term is the reactive coupling one described above.

Once the system has acquired a stationary dynamics we have perturbed both the Master and the Slave with a "kick" that would take the role of thermal forces or other sources of instability impacting the system. We have implemented the "kick" as a sudden change of the deformation of each of the OMCs of up to 10% with respect to the value taken in the previous time step. When the coupled system is initially found in a synchronized state, the phase trajectory described by the Slave after the kick is an oscillatory one around the synchronized state limit cycle solution performed at a frequency of ($\Omega_M - \Omega_S$). The amplitude of this hypermodulation is exponentially decaying with time so that after a few cycles the synchronized state stationary solution is recovered. This effect is better illustrated by plotting the temporal trace associated to $u_S$ after the "kick" (Fig. S6a) but a consistent effect is also experienced by $N_S$ and $\Delta T_S$ since they are coupled to $u_s$.

On the contrary, the Master overdampedly returns to the limit cycle when driven away from it. When comparing Fig. S6a and Fig. S6b it is clear that it takes more time for the Slave to converge again to its stationary limit cycle. After several μs the system has fully recovered the stationary synchronized dynamics, recovering the π phase-shift between the phase trajectories.

Fig. S7 shows the Fourier transform of the simulated optical transmission (RF signal) corresponding to the data showed in Fig. S7, i.e., of temporal traces registered after the "kick". Both RF spectra display a main peak at $\Omega_{sync}$~$\Omega_M$, the clear difference between them being that the Slave spectra display sidebands associated to the modulation showed in Fig. S7a. On the contrary, no sidebands appear on the Master RF spectrum. This is consistent with what observed experimentally, for example in Figs. 2g and 2h.

**Fig. S1.**

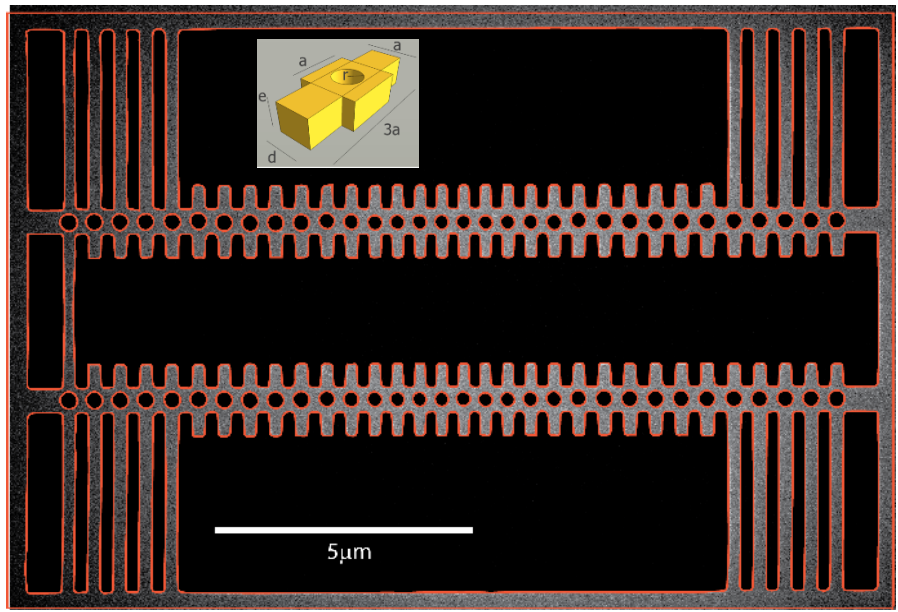

**Fig. S1. Top-view SEM micrograph of the coupled Optomechanical Crystal Cavities.** The extracted geometrical contour imported by the FEM solver is depicted in red. A sketch of the unit-cell is shown in the inset.

**Fig. S2.**

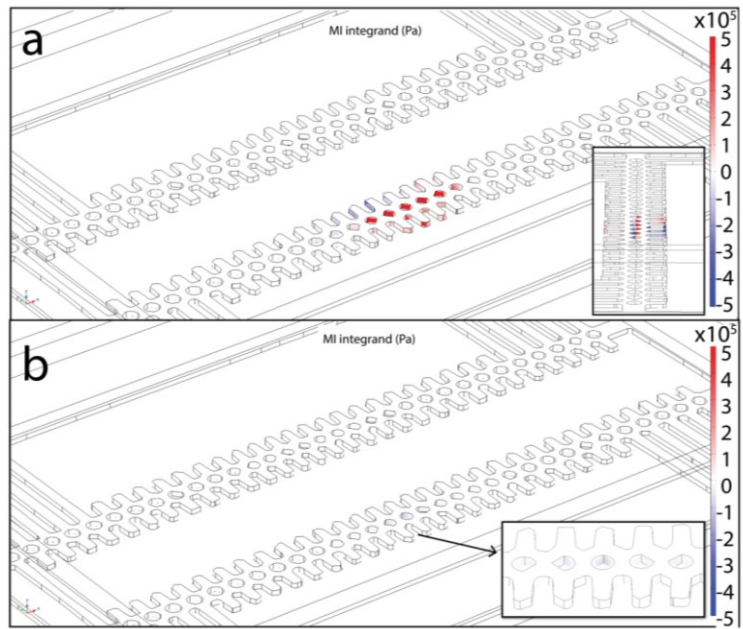

**Fig. S2. Optomechanical coupling contributions.** Normalized surface density of the integrand in Eq. S1 for the Slave optical mode, showing the contributions to $g_{MI}$ of the mechanical mode associated to the Slave (panel a) and the Master (panel b).

**Fig. S3.**

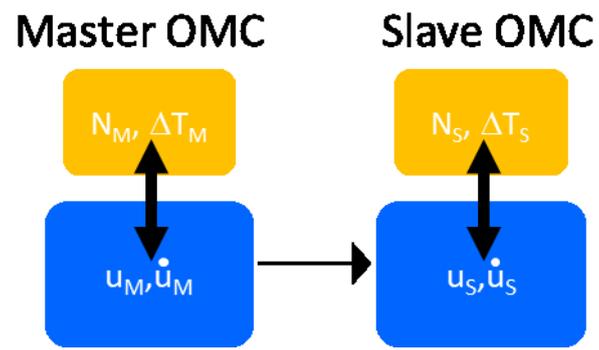

**Fig. S3. Schematic of the modelled system of coupled OMCs.**

**Fig. S4.**

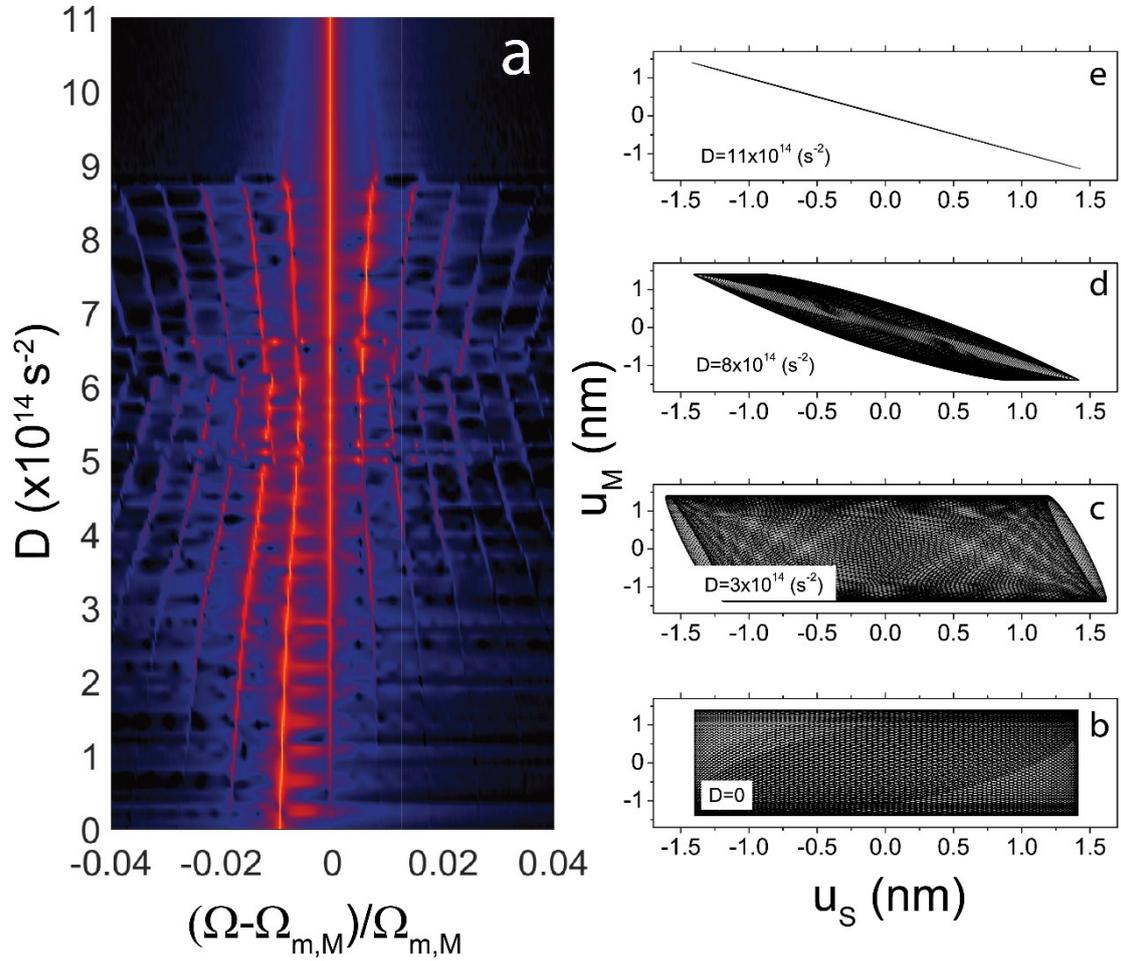

**Fig. S4. Numerical simulations of the coupled OM cavities**. **a**, Colour contour plot of the simulated radio-frequency spectrum of the optical transmission as a function of the coupling constant ($D$). **b-d**, Phase portraits of the deformation of the Master cavity ($u_M$) and the Slave cavity ($u_S$) for different values of $D$.

**Fig. S5.**

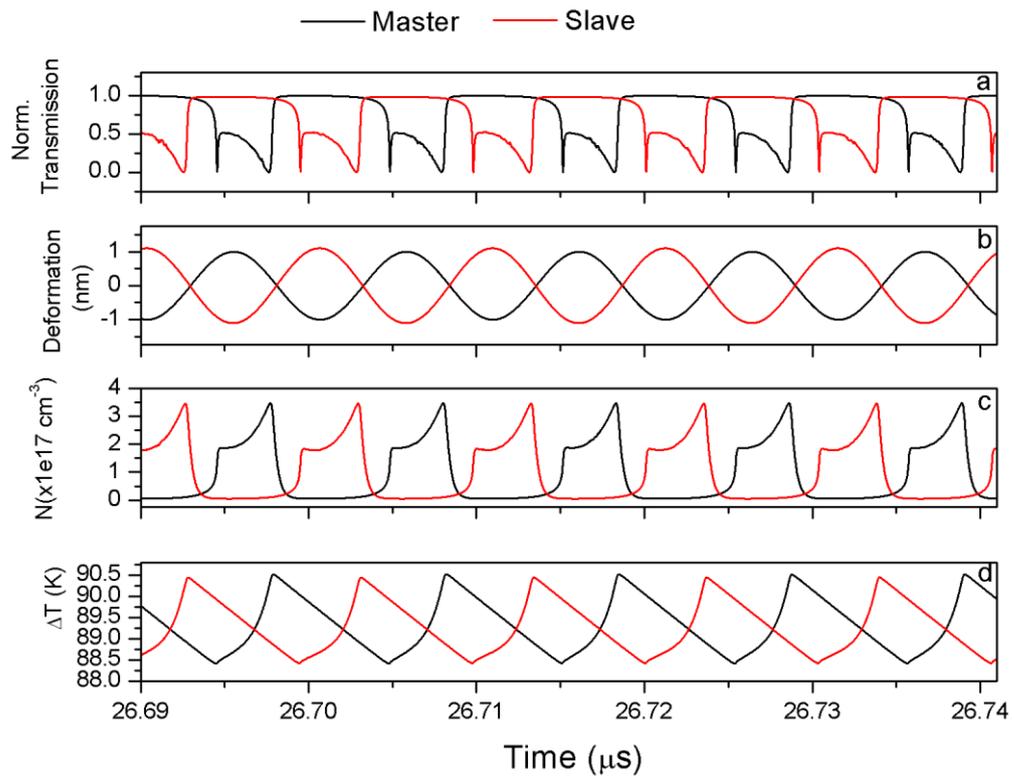

**Fig. S5.** Temporal dynamics of the transmitted optical signal (panel a), deformation (panel b), free-carrier-population (panel c) and the average cavity temperature increase (panel d) associated to the Master (black) and the Slave (red) when the coupled system is in a synchronized state as in Figure 4d.

**Fig. S6.**

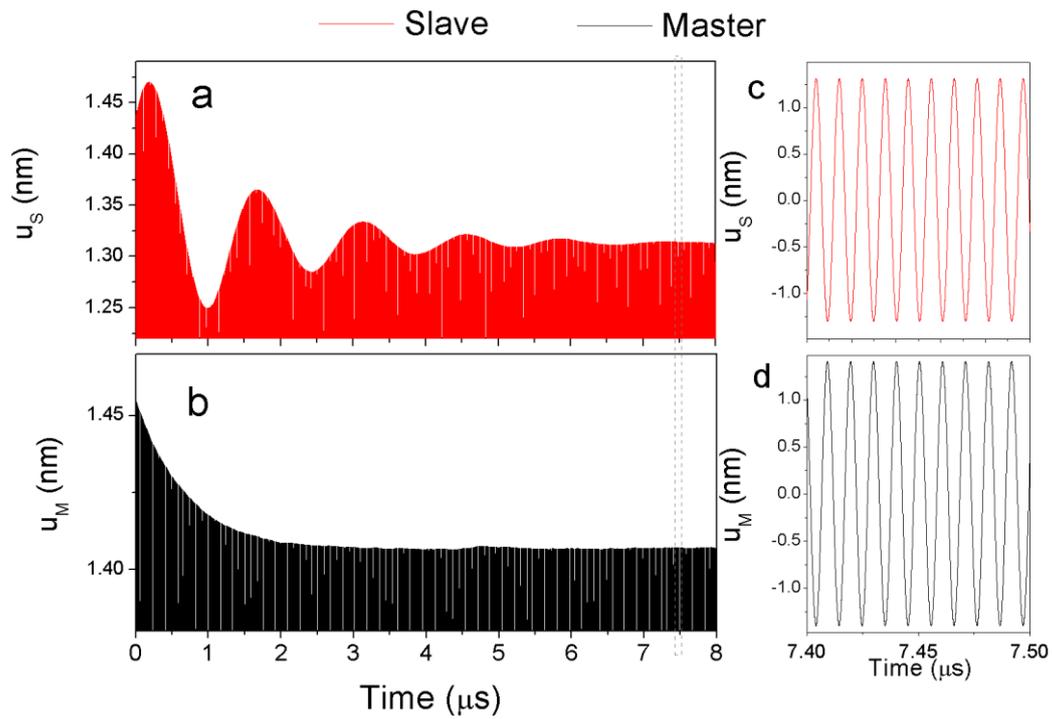

**Fig. S6.** Temporal traces of the mechanical deformation associated to the Slave (panels a and c) and Master (panels b and d) after a "kick" has been applied to the system. A zoom of a region where the stationary regime has been achieved again is reported on the right panels.

**Fig. S7.**

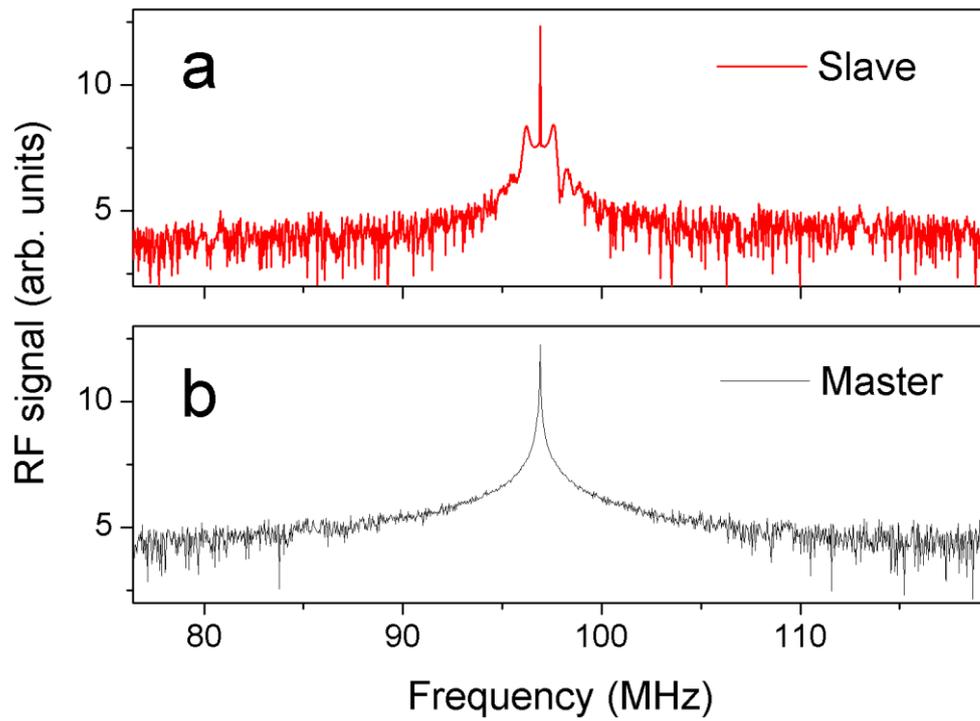

**Fig. S7.** Simulated radio-frequency (RF) spectrum of the optical transmission of the Slave (panel a) and the Master (panel b) when a "kick" has been applied to the system.